\documentclass[letter]{ptptex}
\usepackage[dvips]{graphicx,color}
\usepackage{multirow}
\usepackage{bm}
\usepackage{wrapft}

\def\beq{\begin{equation}}
\def\eeq{\end{equation}}
\def\bea{\begin{eqnarray}}
\def\eea{\end{eqnarray}}

\def\fun#1#2{\lower3.6pt\vbox{\baselineskip0pt\lineskip.9pt
\ialign{$\mathsurround=0pt#1\hfil##\hfil$\crcr#2\crcr\sim\crcr}}}
\notypesetlogo  %comment in if to eliminate PTPTeX logo
\title{%        %You can use \\ for explicit line-break.
Three-Body Model Calculation of Spin Distribution in Two-Nucleon
Transfer Reaction
}

\author{%       %Use \scshape for the family name.
Kazuyuki \textsc{Ogata},$^{1,2}$\thanks{E-mail: ogata@phys.kyushu-u.ac.jp}
Shintaro \textsc{Hashimoto},$^{2}$
and
Satoshi \textsc{Chiba}$^{2}$
}

\inst{%     %Affiliation, neglected when [addenda] or [errata].
$^{1}$Department of Physics, Kyushu University,
Fukuoka 812-8581, Japan
\\
$^2$Advanced Science Research Center, Japan Atomic Energy Agency, \\Ibaraki 319-1195, Japan
}

\abst{%         %This abstract is neglected when [addenda] or [errata].
The differential cross sections of two-nucleon transfer reactions
$^{238}$U($^{18}$O,$^{16}$O)$^{240}$U around 10~MeV per nucleon are
calculated by one-step
Born-approximation with a $^{16}$O$+^{2}n+^{238}$U three-body model.
The three-body wave function in the initial channel is obtained with the
continuum-discretized coupled-channels method,
and that in the final channel is evaluated with adiabatic
approximation.
The resulting cross sections have a peak around the grazing angle,
and the spin distribution, i.e., the cross section at the
peak as a function of the transferred spin, is investigated.
The shape of the spin distribution is found not sensitive to
the incident energies, optical potentials, and treatment of the
breakup channels both in the initial and final states,
while it depends on the excitation energy of the residual
nucleus $^{240}$U. The peak of the spin distribution
moves to the large-spin direction as the excitation energy
increases. To fulfill the condition that the peak position
should not exceeds 10$\hbar$, which is necessary for the
surrogate ratio method to work, it is concluded that
the excitation energy of $^{240}$U must be less than 10~MeV.
}

\begin{document}
\maketitle

Determination of neutron-induced reaction cross sections of unstable
nuclei is one of the most important challenges for nuclear physics and
its application.
Systematic data of fission or neutron capture cross sections in the
neutron-induced reaction with minor actinides (MAs) and long-lived
fission products (LLFPs) are necessary for theoretical designs of the
next generation nuclear plant. \cite{iaea08} \
Such data also play a major role in discussing the nucleosynthesis of
the s- and r-processes in nuclear astrophysics. \cite{arnett,pagel} \
It is, however, difficult to measure these reaction cross sections
directly with currently available experimental techniques;
both neutron and unstable nuclei cannot be used as a target
because of their short lifetime.

The surrogate reaction method
\cite{cram70a,cram70b,back74a,back74b,brit79} is an indirect technique
to obtain the neutron cross section from an analysis of multi-nucleon
transfer reactions or inelastic scatterings producing the same compound
nucleus as that created by the desired (neutron-induced) reaction.
This simplest approach, so-called the surrogate ratio method
(SRM), is based on the Weisskopf-Ewing (WE) approximation, i.e.,
the decay branching ratio of the fission or capture process is assumed
to be independent of the spin-parity $J^\pi$ of the compound nucleus
populated. However, it is very difficult, or almost impossible, to find
a surrogate reaction in which this assumption is satisfied. Recently,
Chiba and Iwamoto \cite{Chiba}
proposed that only a weaker condition, which is called weak
WE condition, was necessary and this idea made the SRM
approach feasible.
The weak WE condition is that a {\it ratio}
of decay rate of a
desired nucleus to that of another one must be constant with respect
to $J^\pi$, which is fulfilled for $J\le10$
(in unit of $\hbar$) in the case of Uranium-isotopes.
Therefore, a gross feature of the spin-parity distribution of the
residual nucleus populated in the surrogate reaction is a key issue
for the SRM approach; the spin-parity distribution
should have a peak somewhat lower than $J=10$.
In previous studies, \cite{esch06,fors07,escher10} \
however, spin-parity distributions of compound
nuclei are assumed rather arbitrarily.

In this Letter, we evaluate the differential cross section of
the $^{238}$U($^{18}$O,$^{16}$O)$^{240}$U reaction at 180~MeV,
with changing the spin transferred to $^{238}$U, and obtain
the spin-parity distribution of $^{240}$U.
We describe the transfer reaction with a $^{16}$O$+^{2}n+^{238}$U
three-body model. We neglect
for simplicity the intrinsic spins of the three particles.
The three-body wave function in the entrance channel is calculated
by the  Continuum-Discretized Coupled-Channels method (CDCC)
\cite{CDCC1,CDCC2,AYK,AKY,Piya}, while that in the exit channel
is described by
adiabatic approximation.~\cite{TJ} \ Thus breakup effects of
both $^{18}$O and $^{240}$U on the spin distribution are investigated.
As for the transition matrix of the
transfer process, we adopt one-step Born approximation (BA).
This approach called CDCC-BA has been applied to many studies
on transfer reactions; see, e.g., Refs.~\citen{S17,Moro}.

In the present study, two neutrons are treated
as a bound particle (di-neutron) and the sequential two-nucleon
transfer process is neglected. Nevertheless, this is the first
calculation of the spin distribution of $^{240}$U populated by
a transfer process by means of a three-body reaction model and
should be regarded as the starting point of the investigation on
the spin distribution associated with the SRM approach.

In the calculation of the $^{16}$O-$^{2}n$ wave function in
CDCC, we include the s-, p-, and d-waves calculated with a
Woods-Saxon potential with the radial (diffuseness) parameter
of $1.27\times (16)^{1/3}$~fm (0.67~fm). We use the separation-energy
method to determine the depth of the potential; we assume the
ground state of the $^{16}$O-$^{2}n$  system is a s-wave state
with the binding energy of 12.2~MeV.
The maximum wave
number of the $^{16}$O-$^{2}n$ continuum is taken to be 1.5~fm$^{-1}$
and the width of the momentum-bin is 0.15~fm$^{-1}$.
As for the distorting potential of $^{2}n$ by $^{238}$U, we take
the neutron global potential of Ref.~\citen{KD} with making the
depth parameters of the real and imaginary parts twice.
The distorting potential between $^{16}$O and $^{238}$U is
evaluated by the double folding model with the
Jeukenne-Lejeune-Mahaux (JLM) nucleon-nucleon interaction~\cite{JLM};
nuclear densities are obtained by Hartree-Fock method
with finite-range Gogny D1S force~\cite{YRS}.
The $^{18}$O-$^{238}$U distorted wave is evaluated up to 30~fm with
an increment of 0.01~fm; we take the number of the partial waves
to be 200.

The wave function of the exit channel is calculated with
adiabatic approximation following Ref.~\citen{TJ}.
As for the binding potential of $^{2}n$ by
$^{238}$U, we adopt a Woods-Saxon form with the radial and
diffuseness parameters of $1.27\times (238)^{1/3}$~fm and 0.67~fm,
respectively. The number of nodes of the $^{2}n$-$^{238}$U wave function
is set to be the same as that of the forbidden states.
Note that the total spin $J$ (in unit of $\hbar$)
of $^{240}$U is equal to the
orbital angular momentum $\ell$ between $^{2}n$ and $^{238}$U
in the present calculation.

The transfer process is described by one-step CDCC-BA as mentioned above.
We made the zero-range
approximation to the transition operator, i.e., the interaction between
$^{2}n$ and $^{16}$O, in the calculation of the transition matrix.
It is confirmed that inclusion of the finite-range correction~\cite{Satchler}
never changes the conclusions below.

%
%%%%%%%%%%%%%%%%%%%%%%%
%%%  Figure 1
%%%%%%%%%%%%%%%%%%%%%%%
%begin{figure}[htbp]
\begin{figure}[t]
\begin{center}
 \includegraphics[width=0.6\textwidth,clip]{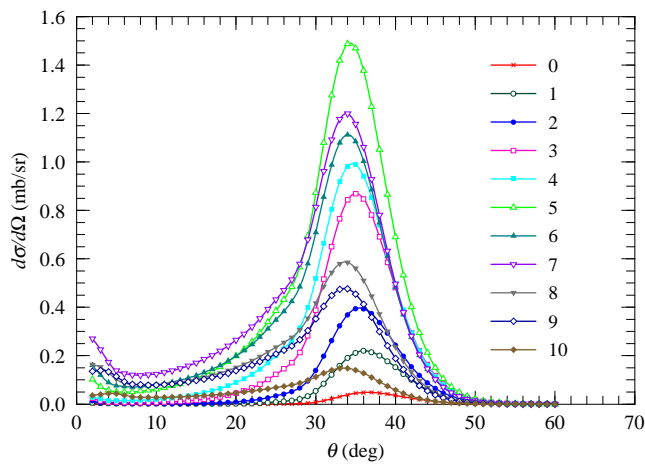}
 \caption{
 (color online) Transfer cross sections of
 $^{238}$U($^{18}$O,$^{16}$O)$^{240}$U at 180~MeV
 for $J=0$--10, as a function
 of the outgoing angle of $^{16}$O in the c.m. frame.
 }
\label{fig1}
\end{center}
\end{figure}
In Fig.~\ref{fig1}, we show the differential cross section of
$^{238}$U($^{18}$O,$^{16}$O)$^{240}$U at 180~MeV (10~MeV per nucleon)
as a function of the outgoing angle $\theta$ of $^{16}$O in the
center-of-mass (c.m.) frame. We put the binding energy $\epsilon_{\rm B}$
between $^{2}n$ and $^{238}$U to be 10.74~MeV that corresponds to
the ground state
of $^{240}$U in the present three-body model. Cross sections for
$0\le J \le 10$ are plotted.
%
%%%%%%%%%%%%%%%%%%%%%%%
%%%  Figure 2
%%%%%%%%%%%%%%%%%%%%%%%
%\begin{figure}[htbp]
\begin{figure}[b]
\begin{center}
 \includegraphics[width=0.5\textwidth,clip]{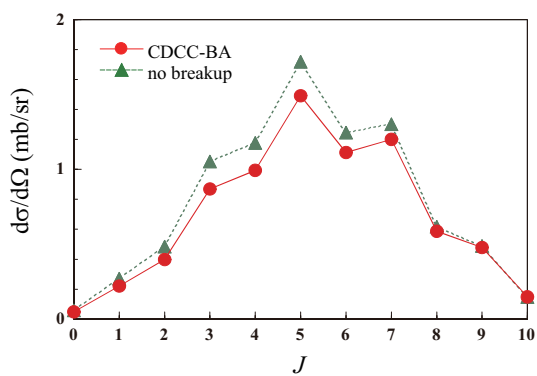}
 \caption{
 (color online) Spin distribution corresponding to the result in
 Fig.~\ref{fig1}. The solid (dashed) line shows the result with
 (without) breakup states of $^{18}$O and $^{240}$U.
 }
\label{fig2}
\end{center}
\end{figure}
They are localized well around
$\theta\sim 35^\circ$. This {\lq\lq}bell-shaped'' angular distribution
is a well-known feature of heavy-ion induced transfer reactions
at energies that are above the Coulomb barrier but low so that
the elastic scattering in each channel is still Fresnel-like.~\cite{Satchler} \
One sees that the peak of the cross section increases with $J$ for
$J\le 5$ and decreases afterwards. This is consistent with
the assumption made in Ref.~\citen{Chiba}.
In Fig.~\ref{fig2}, we show the spin distribution; the solid line is the
result of CDCC-BA and the dashed line is that obtained with neglecting
breakup states of both $^{18}$O and $^{240}$U.
One sees that breakup effects on the spin distribution are
very small;
the total breakup cross section is very small ($\sim 2.8$~mb).
Although they cause change in the absolute values of the spin
distribution by 20\% at most, the gross features of present interest,
i.e., the peak position and the shape, never change.
The renormalization factors of the JLM interaction~\cite{JLM}
are found to have the same effects on the spin distribution.

%
%%%%%%%%%%%%%%%%%%%%%%%
%%%  Figure 3
%%%%%%%%%%%%%%%%%%%%%%%
%\begin{figure}[htbp]
\begin{figure}[t]
\begin{center}
 \includegraphics[width=1.0\textwidth,clip]{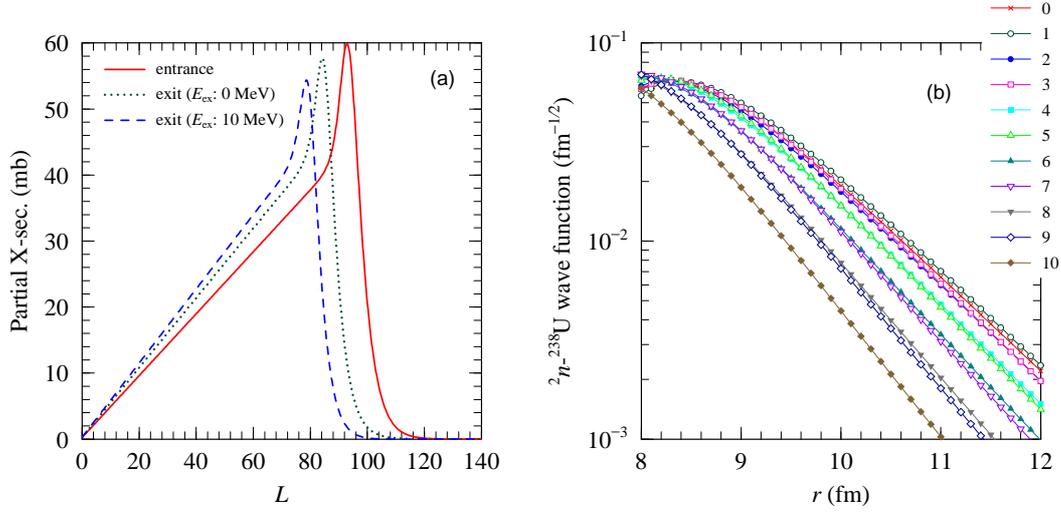}
 \caption{
 (color online) (a) Partial elastic cross sections for the
 entrance (dashed line) and exit (dotted and dashed lines) channels.
 The dotted and dashed lines correspond to
 $\epsilon_{\rm B}=10.74$ and 0.74~MeV, respectively.
 (b) Radial part of the $^{2}n$-$^{238}$U wave function
 multiplied by $r$.
 }
\label{fig3}
\end{center}
\end{figure}
Because of the shortness of the wave number and strong absorption
due to the target nucleus, the transfer process considered in
this study can be interpreted with a simple picture.
In the left panel of Fig.~\ref{fig3}, we show the partial elastic cross
sections (PEX)
for the entrance (solid line) and exit (dotted line) channels.
Each PEX shows a narrow peak at a grazing momentum $L_{\rm g}$.
The difference between the $L_{\rm g}$ for the two channels,
$\Delta L_{\rm g}$,
gives a constraint for $J$, i.e., a transfer process for
$J \sim \Delta L_{\rm g}$
is preferred. On the other hand, as shown in the right panel of
Fig.~\ref{fig3},
the bound state wave function
between $^{2}n$ and $^{238}$U at the grazing radius $r_{\rm g} \sim 10$~fm
decreases with $J$.
Thus, these two features shown in Fig.~\ref{fig3} determine
the spin distribution shown in Fig.~\ref{fig2}.

%
%%%%%%%%%%%%%%%%%%%%%%%
%%%  Figure 4
%%%%%%%%%%%%%%%%%%%%%%%
%\begin{figure}[htbp]
\begin{figure}[t]
\begin{center}
 \includegraphics[width=0.7\textwidth,clip]{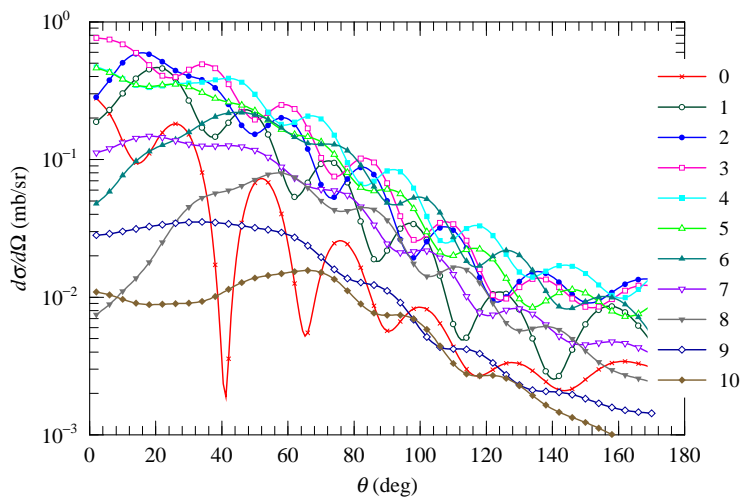}
 \caption{
 (color online) Same as in Fig.~\ref{fig1} but for
 $^{238}$U($^{3}$He,$p$)$^{240}$Np at 30 MeV.
 }
\label{fig4}
\end{center}
\end{figure}
Figures~\ref{fig1}, \ref{fig2}, and \ref{fig3} indicate
some advantages to use
$^{238}$U($^{18}$O,$^{16}$O)$^{240}$U at around 10~MeV per nucleon
as a surrogate reaction. First, the bell-shaped angular distribution
gives a clear criterion for the scattering angle to be measured.
Second, a simple classical picture can be used to interpret the
reaction process. Third, the spin distribution has a peak at
a rather small value of $J$, i.e., $J=5$.
On the other hand, if one uses a ($^{3}$He,$p$) reaction,
the $J$ dependence of the cross section becomes complicated
and
the spin distribution is very sensitive to the detection angle.
This can be seen in Fig.~\ref{fig4}, in which
the cross section of $^{238}$U($^{3}$He,$p$)$^{240}$Np at 30~MeV
(10~MeV per nucleon) calculated with the same framework is shown.

%
%%%%%%%%%%%%%%%%%%%%%%%
%%%  Figure 5
%%%%%%%%%%%%%%%%%%%%%%%
%\begin{figure}[htbp]
\begin{figure}[b]
\begin{center}
 \includegraphics[width=0.5\textwidth,clip]{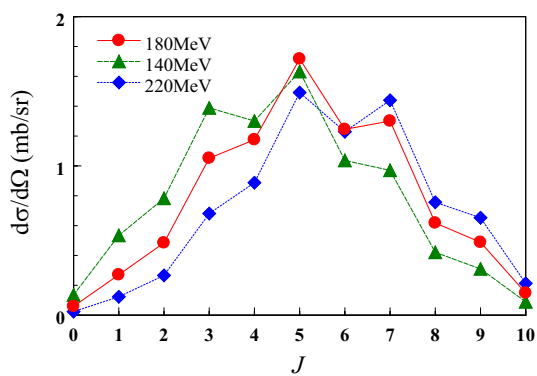}
 \caption{
 (color online) Same as in Fig.~\ref{fig2} but for
 the incident energies of 180~MeV (solid line), 140~MeV (dashed line),
 and 220~MeV (dotted line).
 }
\label{fig5}
\end{center}
\end{figure}
We show in Fig.~\ref{fig5} the energy dependence of the
spin distribution. The solid, dashed, and dotted lines correspond
to $E=180$, 140, and 220~MeV, respectively; breakup effects
of $^{18}$O and $^{240}$U are not included.
Although the peak is located at $J=5$ for all these energies,
the distribution seems to slightly shift to the smaller (larger) $J$
at lower (higher) energies. This suggests that
$^{238}$U($^{18}$O,$^{16}$O)$^{240}$U at lower energies is more suitable
for the SRM. It should be noted, however,
that the incident energy should be larger than the Coulomb barrier
height.

%
%%%%%%%%%%%%%%%%%%%%%%%
%%%  Figure 6
%%%%%%%%%%%%%%%%%%%%%%%
\begin{figure}[htbp]
%\begin{figure}[t]
\begin{center}
 \includegraphics[width=0.5\textwidth,clip]{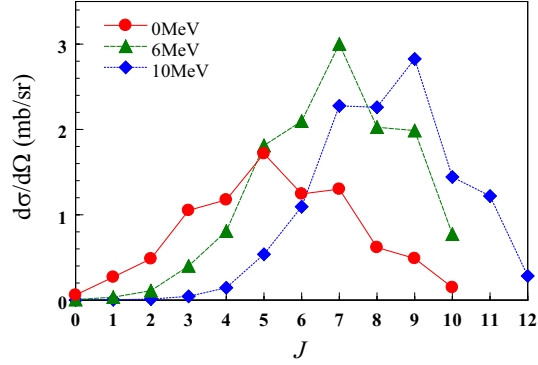}
 \caption{
 (color online)  Same as in Fig.~\ref{fig2} but for
 $E_{\rm ex}=0$~MeV (solid line), 6~MeV (dashed line),
 and 10~MeV (dotted line).
 }
\label{fig6}
\end{center}
\end{figure}
In the calculations shown above, we take $\epsilon_{\rm B}=10.74$~MeV
and the $^{2}n$ transfer process to the ground state of
$^{240}$U is investigated. In the surrogate reaction method, one aims
to produce a compound nucleus $^{240}$U that has energy above
the $n$-$^{239}$U threshold located at 5.93~MeV from the ground
state of $^{240}$U. Thus, we take
$\epsilon_{\rm B}=4.74$ and 0.74~MeV, which correspond to
the excitation energy $E_{\rm ex}$
of $^{240}$U of 6.0 and 10.0~MeV, respectively.
Note that in the present three-body model calculation,
the excited state of $^{240}$U is described by a bound state of
$^{2}n$ by $^{238}$U. For $\epsilon_{\rm B}=0.74$~MeV,
we set the maximum radius between
$^{18}$O and $^{238}$U to be 45~fm in evaluation of
the transition matrix.

Figure~\ref{fig6} shows the spin distribution for
$\epsilon_{\rm B}=4.74$ (dashed line) and 0.74~MeV (dotted line).
Again, we neglect breakup effects of $^{18}$O and $^{240}$U.
Also shown by the solid line is the result for
$\epsilon_{\rm B}=10.74$~MeV, i.e., the dashed line in Fig.~\ref{fig2}.
One sees that the spin distribution somewhat shifts to
the high-$J$ direction as $E_{\rm ex}$ increases.
This is due to the increase in the reaction $Q$-value, hence
$\Delta L_{\rm g}$; see the dashed line in Fig.~\ref{fig3} that is
the PEX in the exit channel corresponding to
$\epsilon_{\rm B}=0.74$~MeV.
It was conjectured in Ref.~\citen{Chiba} that the SRM
worked well unless a compound nucleus with $J > 10$
was populated by a surrogate reaction. Thus, we conclude from
Fig.~\ref{fig6} that $E_{\rm ex}=10$~MeV is the upper limit of
the SRM.

It should be noted that
unlike schematic spin distributions of Escher in Ref.~\citen{escher10},
the spin distribution of Fig.~\ref{fig6} is skewed so that low $J$-values
are enhanced.
Therefore, contributions for $J\ge10$ are not the major part of the
populated compound nuclei even in these heavy nuclear systems.
Another remark is that we must multiply the above result by the
level-density of the final state to evaluate the population cross section
of different $J^\pi$ states.
The $J$-dependence of the level-density calculated with the Fermi gas
formula using the deformation parameter and mass of $^{240}$U shown in
Ref.~\citen{KTUY} is very weak in the region of $J\le10$.
This multiplication, therefore, gives no change in the conclusion above.

%
%%%%%%%%%%%%%%%%%%%%%%%
%%%  Figure 7
%%%%%%%%%%%%%%%%%%%%%%%
\begin{figure}[htbp]
%\begin{figure}[t]
\begin{center}
 \includegraphics[width=0.5\textwidth,clip]{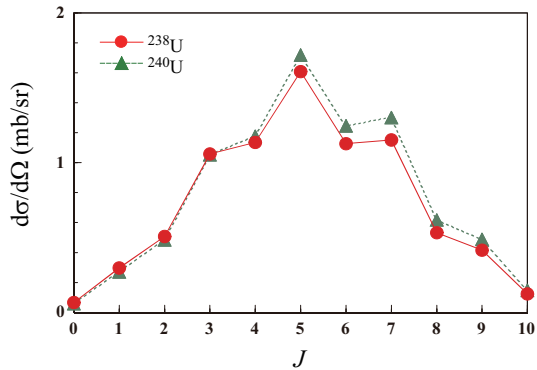}
 \caption{
 (color online) Same as in Fig.~\ref{fig2} but for
 $^{236}$U($^{18}$O,$^{16}$O)$^{238}$U at 180~MeV (solid line).
 Result for $^{238}$U($^{18}$O,$^{16}$O)$^{240}$U at 180~MeV
 is also shown by the dashed line for comparison.
 }
\label{fig7}
\end{center}
\end{figure}
An important point of the SRM proposed in Ref.~\citen{Chiba} is
to find a pair of surrogate reactions, the spin distributions of
which are equivalent (for $J \le 10$). We show in Fig.~\ref{fig7}
the comparison of the spin distributions for the
$^{238}$U($^{18}$O,$^{16}$O)$^{240}$U (solid line) and
$^{236}$U($^{18}$O,$^{16}$O)$^{238}$U (dashed line)
reactions at 180~MeV;
each of the residual nuclei is assumed to be a ground state.
One clearly sees that the two results show good agreement.
Therefore, the two reactions can be used as a pair of surrogate
reactions required in the SRM in Ref.~\citen{Chiba}.
It should be noted that
we ignore the spectroscopic factors of the populated states
in the present calculation. This can be justified by the fact that
in the SRM two spectroscopic factors corresponding to two surrogate
reactions are expected to cancel out by taking the ratio of
the spin distributions; these two spectroscopic factors
are considered to have similar $J^\pi$-dependence.

In summary, we present the differential cross sections of the
two-nucleon transfer reaction $^{238}$U($^{18}$O,$^{16}$O)$^{240}$U and
the spin distribution of the residue calculated with CDCC-BA based on
a $^{16}$O$+^{2}n+^{238}$U three-body model.
The angular distribution of the transfer reaction at 180~MeV with the
binding energy $\epsilon_{\rm B}=10.74$~MeV, which  corresponds to the
ground state of $^{240}$U, is bell-shaped with the peak around the outgoing
angle $\theta\sim 35^\circ$, while that of
$^{238}$U($^{3}$He,$p$)$^{240}$Np at 30~MeV is very complicated.
The spin distribution of the $^{238}$U($^{18}$O,$^{16}$O)$^{240}$U cross
section at the peak is explained by the difference between the grazing
momenta in the initial and final channels, and has the peak at $J=5$.
The shape of the spin distribution hardly depends on the incident energy
and the optical potential parameters.
The breakup cross section of $^{18}$O in the present system is very
small and the breakup effect on the gross features of the spin distribution
is negligible.
With increasing the excitation energy of the $^{240}$U produced, the
peak of the spin distribution shifts to the larger $J$.
Because the peak for $\epsilon_{\rm B}=0.74$~MeV
corresponding to the neutron energy of 4.07~MeV is located at
$J\sim9$, this energy will be
the upper limit that the SRM works well, according to the
conclusion in Ref.~\citen{Chiba}.

For more detailed studies on the spin distribution of two-neutron transfer
reaction,
one should take into account the contribution of the sequential transfer
of the two neutrons by a two-step process. This requires three-body
description of $^{18}$O and $^{240}$U, i.e., four-body CDCC.
Effects of the deformation of $^{238}$U, which produces rotational states,
will also be important. It is expected that the
spin distribution spreads slightly because of the coupling with the
rotational angular momentum, which is suggested to be around $2\hbar$ by
the recent work with a dynamical model based on multi-dimensional
Langevin equations in Ref.~\citen{Aritomo}.

\vspace{3mm}

We would like to thank K.~Hagino for helpful discussions.
We also acknowledge valuable suggestions from the experimental viewpoint
with K.~Nishio.
The present study is a result of \lq\lq Development of a Novel Technique
for Measurement of Nuclear Data Influencing the Design of Advanced Fast
Reactors'' entrusted to Japan Atomic Energy Agency (JAEA) by the
Ministry of Education, Culture, Sports, Science and Technology of Japan
(MEXT).

%%--------------------------------------------------------------------%%
%%                           References                               %%
%%--------------------------------------------------------------------%%

\end{document}